\documentclass{aa}

\usepackage{amsmath}
\usepackage{graphicx}
\usepackage{txfonts}
\usepackage{xcolor}
\graphicspath{{figures_aa/}}
\definecolor{citationblue}{RGB}{0,70,150}
\let\aaoriginalcitep\citep
\renewcommand{\citep}[2][]{%
  \textcolor{citationblue}{\aaoriginalcitep[#1]{#2}}%
}
\let\aaoriginalcitet\citet
\renewcommand{\citet}[2][]{%
  \textcolor{citationblue}{\aaoriginalcitet[#1]{#2}}%
}

\newcommand{\sii}{[S\,{\sc ii}]}

\newcommand{\nev}{[Ne\,{\sc v}]}
\newcommand{\heii}{He\,{\sc ii}}

\newcommand{\msun}{M_\odot}

\begin{document}

\title{Does galaxy-scale compactness regulate nebular electron density? A cross-scale test with DESI}

\author{Shihong Liu\inst{1,2}\and Yu Rong\inst{1,2}\corrauth{rongyua@ustc.edu.cn}}
\authorrunning{Liu \& Rong}
\titlerunning{Compactness and nebular density}

\institute{Department of Astronomy, University of Science and Technology of
China, Hefei, Anhui 230026, China\\
\and School of Astronomy and Space Sciences, University of Science and
Technology of China, Hefei 230026, China}

\date{}

\abstract
{Nebular electron density, $n_e$, measured from the
\sii$\lambda6716/\lambda6731$ doublet traces ionized gas in small-scale
H\,{\sc ii} regions and diffuse ionized structures, whereas galaxy stellar
surface density, $\Sigma_\star$, measures compactness on kpc scales. Whether
these quantities are correlated is therefore a cross-scale question rather
than a geometrical identity.}
{We test whether nebular electron density varies systematically with
galaxy-scale stellar compactness after controlling for the correlated effects
of stellar mass and star-formation rate (SFR).}
{We use Dark Energy Spectroscopic Instrument Data Release 1 spectra to test
this cross-scale connection in AGN-clean star-forming galaxies at
$0.10<z<0.44$.  We divide each redshift interval into tertiles of
$\Sigma_\star=M_\star/(2\pi R_e^2)$. We construct one controlled sample in
which the $M_\star$ distributions are matched across the three compactness
tertiles, and a second in which their joint $M_\star$--SFR distributions are
matched, thereby minimizing the influence of these correlated galaxy
properties. For each compactness tertile, we stack the galaxy spectra with
equal weight and use the stacked spectrum to measure the average electron
density of the galaxies in that sample. No \sii\ line-flux or S/N cut is
imposed on the parent sample.}
{The average electron
density increases systematically toward higher stellar surface density in all
redshift intervals. At $T_e=10^4$ K, the
high-minus-low compactness density differences are
$47.5^{+2.5}_{-2.6}$, $63.7^{+2.7}_{-2.8}$, and
$57.8^{+3.7}_{-2.7}$ cm$^{-3}$ in the $M_\star$-matched samples, and
$38.0^{+1.9}_{-2.4}$, $56.1^{+2.1}_{-2.2}$, and
$46.1^{+3.2}_{-4.6}$ cm$^{-3}$ after also matching SFR. 
}
{This is the first controlled population-level evidence for a correlation
between kpc-scale galaxy stellar surface density (compactness) and the average
electron density of ionized gas on subgalactic scales. Its persistence after
matching both $M_\star$ and SFR suggests that a more compact stellar
configuration may promote higher gas pressure and/or stronger confinement of
ionized gas, linking galaxy-scale structure to nebular conditions.}

\maketitle

\section{Introduction}

Nebular electron density, $n_e$, is a basic property of the ionized
interstellar medium.  It controls collisional de-excitation, enters
photoionization modelling and abundance calibrations, and traces the thermal
pressure of line-emitting gas \citep{osterbrock2006,kewley2002,dopita2006}.
The forbidden \sii\ $\lambda\lambda6716,6731$ doublet is a standard density
diagnostic whose calibration and density limits have been discussed in detail
for nebular spectra \citep{proxauf2014}. In a fibre spectrum it does not resolve individual H\,{\sc ii}
regions; instead, it measures a density diagnostic for the nebular structures
contained within the aperture. Measurements in local and distant galaxies have established
that nebular density varies with redshift and correlates with global
quantities such as stellar mass, SFR, and sSFR
\citep{brinchmann2008,shirazi2014,sanders2016,kaasinen2017,
kaasinen2018,Isobe2023,Topping2025,Reddy2023,liurong2026,
ferruit2022,deGraaff2024,heintz2024}.  Recent JWST studies have extended
such measurements to the early Universe, including both individual and
stacked spectra, and have shown that the inferred density depends on the
line-detection threshold and the population represented by the measurement.
These diagnostics can be strongly affected by line-bright H\,{\sc ii} regions,
diffuse ionized gas, and aperture coverage.

Stacking provides a way to recover this population-level quantity when the
density-sensitive doublet is too weak in individual spectra.  This approach
has recently been demonstrated with stacked JWST/NIRSpec spectra of ordinary
low-mass galaxies at $z\simeq2$--7, yielding [S\,{\sc ii}]-based densities
below the individual-doublet detection threshold and showing that individually
measurable strong-line galaxies can be biased toward higher densities
\citep{liurong2026}.

This motivates a different question.  Galaxy surface density (compactness) is commonly used to
describe the stellar distribution, central mass build-up, and gas evolution
\citep{kauffmann2003b,fang2013,barro2013,woo2015}, but it is measured
on kpc scales.  In contrast, a \sii\ density traces the conditions in
H\,{\sc ii} regions and diffuse ionized structures on much smaller scales.
The key issue is therefore not whether a compact galaxy is denser in a purely
geometrical sense, but whether its global structure leaves a measurable
imprint on the density of the gas that emits the nebular lines.

There are reasons to expect such an imprint, and equally good reasons to doubt
it.  A compact stellar body may produce a deeper potential, a larger gas
column, and higher mid-plane pressure, thereby changing the pressure and
ionization structure of its star-forming regions
\citep{elmegreen1989,blitz2006,leroy2008}.  Conversely, diffuse galaxies can
contain substantial H\,{\sc i} reservoirs \citep{huang2012,saintonge2017},
and the line-emitting gas may be distributed among H\,{\sc ii} regions,
diffuse ionized gas, and regions sampled at different physical impact
parameters.  Gas fraction, feedback, metallicity, aperture coverage, and the
relative contributions of different ionized structures can therefore decouple a local line
ratio from global stellar structure \citep{saintonge2017,belfiore2016,
zhang2017,lacerda2018}.  A kpc-scale stellar density should consequently not
be assumed to map one-to-one onto a pc- or sub-kpc-scale gas density.

The physical link is also not expected to be one-dimensional.  At fixed
$M_\star$ and SFR, a smaller effective radius may increase gas pressure and
the pressure of star-forming regions, but changes in the mixture of
H\,{\sc ii} regions and diffuse gas may erase the relation.  Existing work
has established important relations between nebular density and redshift,
stellar mass, and star formation \citep{brinchmann2008,shirazi2014,
sanders2016,kaasinen2017,kaasinen2018}, and separately between
stellar compactness and galaxy structure, gas content, and evolution
\citep{kauffmann2003b,fang2013,barro2013,woo2015,zhang25}.  To our knowledge,
however, no previous population-level study has directly tested the relation
between nebular electron density and galaxy stellar surface density while
controlling their shared dependences on $M_\star$ and SFR. This is the specific
gap addressed here. We refer to this cross-scale correlation as the
$n_e$--$\Sigma_\star$ relation. DESI makes
the experiment possible through the signal-to-noise gain of median stacking:
individual galaxies need not yield a reliable doublet ratio for the parent
population to be tested.

This framing leads to a concrete prediction. If galaxy stellar surface density
is physically linked to the conditions of the ionized interstellar medium,
then the \sii\ doublet ratio should vary systematically with
$\Sigma_\star$ within narrow redshift intervals, and the ordering should
survive matching in $M_\star$.  If the connection is mediated entirely by the
global star-formation budget, it should largely disappear after matching both
$M_\star$ and SFR.  The two controlled stacks therefore distinguish a
structural contribution from a correlation inherited from the usual global
galaxy scaling relations.

\section{Data and sample selection}

We use DESI Data Release 1 coadded spectra \citep{desispectrograph,desidr1}
and the DESI stellar-mass and emission-line value-added catalogue of
\citet{zou2024}.  The stellar masses, $M_\star$, used throughout this work are
taken from this catalogue, which also provides the SFRs and emission-line
measurements used for sample selection, matching, and AGN rejection.  We
select main-survey galaxies with official DESI redshifts
$0.10<z<0.44$ (for $z>0.44$, [S II] lines are out of the wavelength range), \texttt{ZWARN}=0, spectral type \texttt{GALAXY},
positive SFR, positive Legacy Surveys
\texttt{SHAPE\_R} \citep{dey2019}, and specific SFR,
$\log({\rm sSFR/yr^{-1}})>-11$ (i.e., star-forming sample).  We require
$\log(M_\star/\msun)\geq8.5$.  Neither \sii\ component flux nor its S/N is
used in selecting the parent sample.

The size information is taken from the Legacy Surveys Tractor morphology
parameters included in the DESI value-added catalogue of \citet{zou2024}.
Specifically, we use the catalogue field \texttt{SHAPE\_R} in the
\texttt{EMLINES\_MASS} table extension.  This is the angular effective-radius
parameter reported by the imaging catalogue, in arcsec; it is not a radius
derived from the DESI spectrum.  We convert \texttt{SHAPE\_R} from arcsec to
physical effective radius using
the flat Planck 2018 cosmology \citep{planck2020}, with
$H_0=67.66$ km s$^{-1}$ Mpc$^{-1}$ and $\Omega_m=0.30966$ as implemented by
the \textsc{Astropy} Planck18 realization.  Thus every $R_e$ is in physical
kpc.  Combining these radii with the catalogue stellar masses from
\citet{zou2024}, we define the stellar surface density, which characterizes
galactic compactness, as
\begin{equation}
\Sigma_\star=\frac{M_\star}{2\pi R_e^2},
\end{equation}
where $\Sigma_\star$ is expressed in $\msun\,{\rm kpc}^{-2}$.

We remove possible AGN and shocks using the NII and \sii\ BPT diagnostics
where the requisite lines have S/N$>3$ \citep{baldwin1981,kauffmann2003,
kewley2001,kewley2006}.  We additionally remove every object with catalogue
\heii$\lambda4686$, \nev$\lambda3346$, or \nev$\lambda3426$ detected at
S/N$>1$.  This conservative veto reduces contamination by hard ionizing
sources without imposing a \sii\ selection.  It is important, however, that
this procedure is most effective for objects with sufficiently strong lines:
for weak-line galaxies, the absence of a significant diagnostic line means
that a weak AGN component cannot always be identified and removed reliably.
This limitation is relevant to the interpretation of an individual object,
but is unlikely to dominate the present population measurement.  Our parent
sample is restricted to low-redshift, actively star-forming systems, for which
the residual contribution of AGN after the standard emission-line
classification is expected to be small compared with the star-forming
population as a whole \citep{kauffmann2003,brinchmann2004}.  We therefore expect a modest
number of incompletely identified AGN to add scatter rather than determine the
observed ordering with $\Sigma_\star$, while retaining this residual
contamination as a caveat.

\begin{figure*}
\centering
\includegraphics[width=0.98\textwidth]{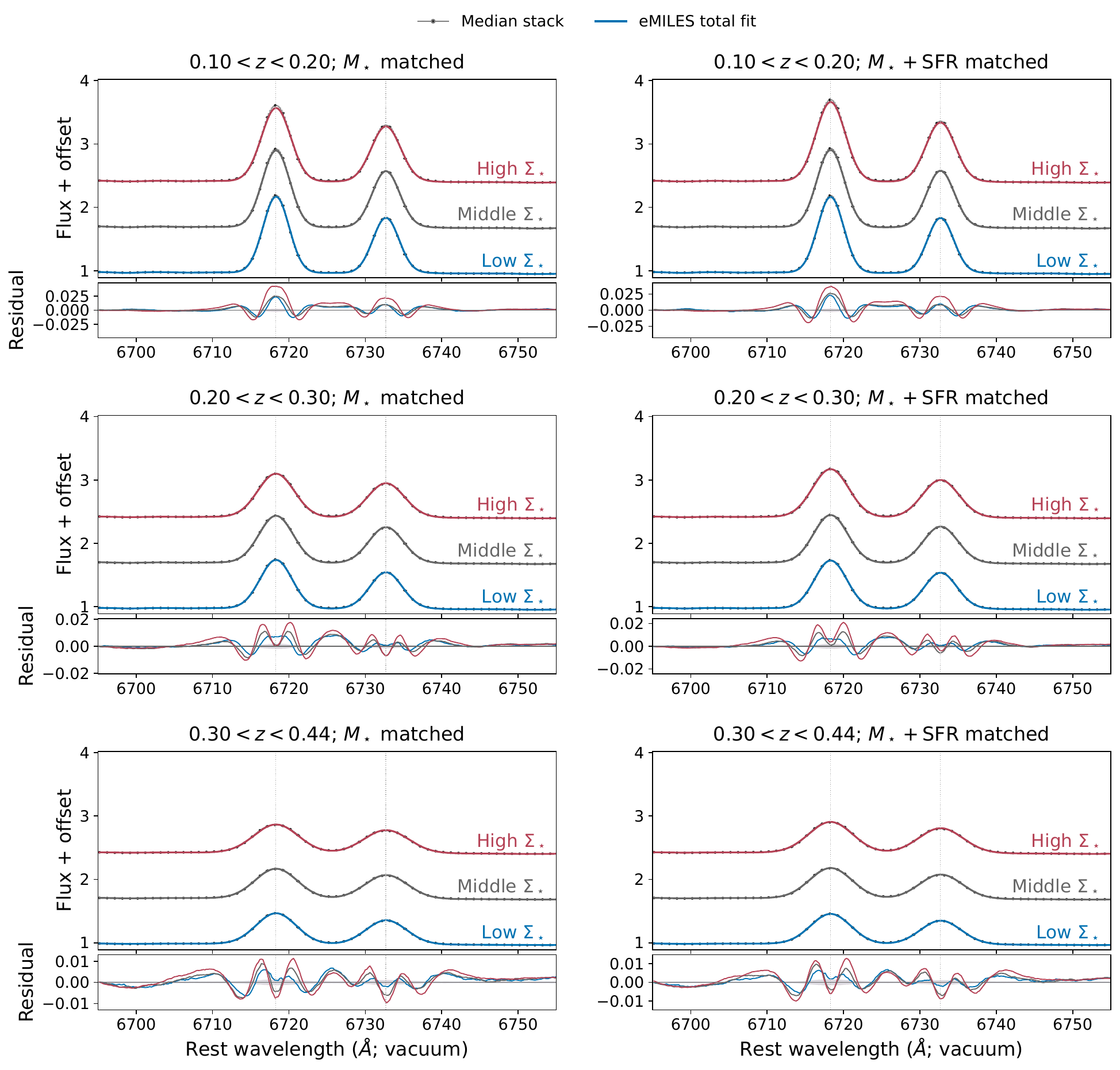}
\caption{Full-sample continuum-normalized median \sii\ spectra and their
eMILES-based fits. In each panel, the grey points and thin grey curves show
the observed stacked spectra obtained by giving every galaxy one equal weight.
The blue, grey, and red curves show the corresponding eMILES total best fits
for the low-, middle-, and high-$\Sigma_\star$ tertiles, respectively; each
total model is the fitted stellar continuum plus the two-Gaussian [S\,{\sc ii}]
doublet model. The lower panels show the residuals defined as the stacked
spectrum minus its total fit. Rows show redshift intervals and columns show
the two matching schemes. The bootstrap error bars are very small because
each stacked spectrum contains a very large number of galaxies.}
\label{fig:stacks}
\end{figure*}

\section{Stacking and measurement}

The scope of this work is to study the relationship between galaxy-scale compactness and nebular electron density. Therefore, individual-galaxy density measurements are unsuitable for the present
experiment, since most galaxies in the parent sample have weak emission lines, so
the [S II] doublet has insufficient S/N for a reliable electron density ($n_e$) measurement in
an individual spectrum.  The galaxies with sufficiently strong lines form a
strongly selected subset: they are preferentially line-bright and have higher
surface brightness, and their completeness can vary with redshift, surface
brightness, and other galaxy properties.  A relation measured only from this
subset could therefore be biased toward the conditions of the brightest
nebular regions and need not describe the parent population.  We instead
retain the weak-line galaxies, divide the sample into bins of
$\Sigma_\star$, and stack the spectra in each bin to measure the
population-level [S II] density relation.  This approach is also supported by
recent JWST work showing that strong-line individual-galaxy samples can differ
from densities recovered from stacked spectra \citep{liurong2026}.

\begin{figure*}
\centering
\includegraphics[width=0.98\textwidth]{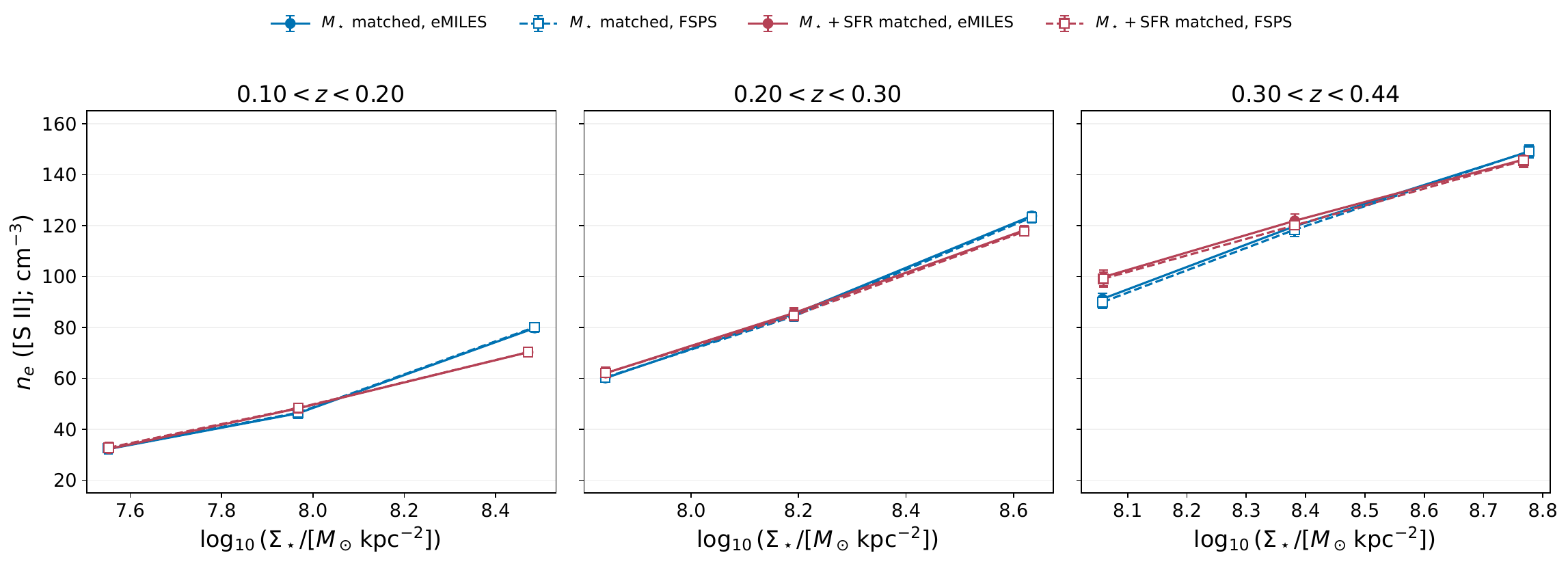}
\caption{Electron density from the \sii\ ratio at $T_e=10^4$ K as a function
of stellar compactness. Blue and red show $M_\star$ matching and
$M_\star+$SFR matching, respectively. Circles with solid lines show the
eMILES measurements, while squares with dashed lines show the FSPS
measurements. The bootstrap error bars are very small because each measurement
is based on a stack containing a very large number of galaxies.}
\label{fig:density}
\end{figure*}

We divide the galaxy sample into three redshift intervals, $0.10<z<0.20$,
$0.20<z<0.30$, and $0.30<z<0.44$.  Within each interval, the AGN-clean
parent population is divided into tertiles of $\log\Sigma_\star$. Stellar mass is a primary parameter governing many galaxy properties,
such as stellar populations, star-formation activity, and gas content
\citep{kauffmann2003b,brinchmann2004,catinella2018}.  We therefore match the
$M_\star$ distributions across compactness tertiles to test the
$n_e$--$\Sigma_\star$ relation at fixed stellar mass. The matching is performed by random, without-replacement sampling within common
property cells.  For the $M_\star$-matched sample, we divide each redshift
interval into cells of width $\Delta z=0.01$ and
$\Delta\log M_\star=0.1$.  In every cell, the three compactness tertiles
define three candidate pools.  If their sizes are $N_1$, $N_2$, and $N_3$,
we draw $N_{\rm min}=\min(N_1,N_2,N_3)$ objects without replacement from
each pool.  Thus the same number of objects is retained from all three
compactness tertiles in every populated cell, while retaining the largest
possible common subsample under the adopted cell definition.

Except for stellar mass, quiescent and star-forming galaxies also have systematically different
compactness distributions \citep{barro2013,fang2013,woo2015}, suggesting
that star-formation activity may also influence the observed
$n_e$--$\Sigma_\star$ relation.  We therefore also perform another matching method to the three compactness tertiles, i.e., matching both
$M_\star$ and SFR, to assess this possible dependence within our
star-forming sample. The $M_\star+$SFR-matched sample uses the identical procedure after further
subdividing every cell into $\Delta\log {\rm SFR}=0.2$ intervals.  

The random
draws use a fixed seed, and each target is used at most once.  Consequently,
the resulting samples have the same binned redshift and $M_\star$
distributions, and the second sample additionally has the same binned SFR
distribution, across the three $\Sigma_\star$ tertiles.  These controls are
performed independently in each redshift interval. 
For each compactness tertile in a redshift bin, we stack the spectra of the selected galaxies in this tertile after matching.  Each galaxy spectrum is firstly corrected for Milky Way extinction using the fibre-map
reddening with the extinction law of \citet{cardelli1989} and the
recalibration of \citet{schlafly2011}, shifted to the rest frame, and interpolated on a
0.2 \AA\ grid
over 6000--6800 \AA. Each spectrum is normalized by the median of the median
fluxes in the fixed continuum windows 6005--6250 and 6760--6790 \AA.
Both windows avoid the principal nebular emission lines and are the same
for all redshifts. The wider first window improves the stability of the
continuum scale for individual weak spectra. This follows the
continuum-normalized stacking approach of \citet{liurong2026}.
A key step is matching the spectral resolution before stacking. For each
galaxy, we use the DESI resolution matrix at the observed wavelength of
\sii\ to estimate the instrumental Gaussian width
$\sigma_{\rm inst,obs}$ and convert it to the rest frame as
$\sigma_{\rm inst,rest}=\sigma_{\rm inst,obs}/(1+z)$. Following the
resolution-matching procedure of \citet{liurong2026}, we then convolve the
rest-frame spectrum with a Gaussian of width
\begin{equation}
 \sigma_{\rm conv}=\left(\sigma_{\rm targ}^{2}
 -\sigma_{\rm inst,rest}^{2}\right)^{1/2},
\end{equation}
where $\sigma_{\rm targ}=1.2$ \AA\ is broader than the native rest-frame
widths of the accepted spectra at the doublet. This operation only smooths
each spectrum to the same poorer local resolution; no deconvolution is
applied. 

The stack is a pixel-wise median in which each normalized galaxy has equal
weight. The density inferred from each stack is therefore the statistical
average $n_e$
of the galaxies in that $(z,\Sigma_\star)$ bin. We estimate
uncertainties with 100 galaxy-level bootstrap realisations.
Within each redshift and compactness bin, galaxies are drawn with replacement
from the fixed matched sample while retaining the original sample size; the
same resampling indices are used at every wavelength. Each realisation repeats
the pixel-wise median stack over 6000--6800~\AA, the stellar-continuum
fit, the doublet fit, and the conversion to $n_e$. 
The bootstrap
realisations are used to estimate uncertainty: 
the 16th--84th percentiles of the 100 realisations give the quoted intervals,
and their pixel-wise dispersion gives the spectral error bars in
Fig.~\ref{fig:stacks}.

Following the continuum-subtraction strategy of \citet{liurong2026}, we fit
the stacked spectra with \textsc{pPXF}
\citep{cappellari2004,cappellari2017}. We use the official \textsc{pPXF}
E-MILES single-stellar-population (SSP) file
\texttt{spectra\_emiles\_9.0.npz}, available from the
\textsc{pPXF} stellar population synthesis (SPS) model repository.\footnote{\url{https://github.com/micappe/ppxf_data}}
This E-MILES subset \citep{vazdekis2016,girardi2000,salpeter1955} adopts
Padova isochrones and a Salpeter initial mass function. We retain 24 age bins
spanning 0.0631--12.59 Gyr
and six metallicities, $[\mathrm{M}/\mathrm{H}]=-1.71,-1.31,-0.71,-0.40,0.00$,
and $+0.22$, giving 144 SSP templates in total.
The templates are restricted to 5900--6900 \AA, normalized over
6005--6250 \AA, logarithmically rebinned to the velocity scale of each stack,
and broadened to the adopted Gaussian instrumental resolution. We fit the
available 6005--6790 \AA\ rest-frame interval without regularization, masking
the principal nebular lines ([O I], H$\alpha$, [N II], He I, and [S II]) and
including an eighth-order additive polynomial. We then subtract the fitted
stellar continuum and fit the [S II] residual over 6695--6755 \AA\ with two
Gaussian components. The Gaussians share a wavelength
shift and width and have independent non-negative integrated fluxes.
Figure~\ref{fig:stacks} shows the total best-fit models on top of the measured
stacks. The fitted integrated flux ratio $R_{\rm SII}=F(\lambda6716)/F(\lambda6731)$
is converted to $n_e$ at $T_e=10^4$ K with \textsc{PyNeb}
\citep{luridiana2015}. For every bootstrap spectrum, the stellar continuum
and doublet are refitted before the ratio is converted to density. This
propagates finite-sample fluctuations through both fitting stages while
leaving the unresampled measurement as the central value.

\section{Results}

Figure~\ref{fig:stacks} shows the median spectra before presenting the derived
density relation.  The two components of [S II] doublet are resolved in every redshift
interval after the common-LSF convolution.  The line-shape change is subtle,
so the three compactness tertiles are vertically offset.  The high-compactness
stack has a smaller $\lambda6716/\lambda6731$ ratio, indicating higher density.

Figure~\ref{fig:density} quantifies the spectral difference.  In the
$M_\star$-matched sample, the electron densities inferred from the low/middle/high
compactness median stacks are
$(32.2^{+1.9}_{-1.9},\,46.3^{+1.2}_{-1.9},\,79.7^{+1.5}_{-1.6})$,
$(60.1^{+1.5}_{-1.6},\,85.3^{+1.9}_{-1.9},\,123.8^{+1.5}_{-2.0})$, and
$(91.2^{+2.3}_{-2.5},\,119.7^{+2.6}_{-2.6},\,149.0^{+2.6}_{-2.4})$
cm$^{-3}$ in the three redshift intervals, respectively.  The corresponding
high-minus-low differences
are $47.5^{+2.5}_{-2.6}$, $63.7^{+2.7}_{-2.8}$, and
$57.8^{+3.7}_{-2.7}$ cm$^{-3}$, respectively. After SFR matching, the corresponding
sequences are $(32.3^{+2.0}_{-1.0},\,48.2^{+1.3}_{-1.3},\,70.3^{+1.3}_{-1.2})$,
$(62.0^{+2.3}_{-1.8},\,85.7^{+2.0}_{-1.9},\,118.2^{+1.7}_{-1.8})$, and
$(99.8^{+2.6}_{-3.3},\,121.9^{+2.7}_{-2.7},\,145.9^{+2.0}_{-2.7})$
cm$^{-3}$, with paired high-minus-low differences of
$38.0^{+1.9}_{-2.4}$, $56.1^{+2.1}_{-2.2}$, and
$46.1^{+3.2}_{-4.6}$ cm$^{-3}$, respectively. The uncertainties on the
individual values are the 16th--84th percentiles of the corresponding
bootstrap distributions; those on the differences are the percentiles of
the bootstrap high-minus-low distributions.  Thus in all redshift bins, $n_e$
increases across the three compactness tertiles.  The relation
persists after controlling the catalogue SFR as well as stellar mass.
Table~\ref{tab:measurements} gives the individual stack values and
galaxy-bootstrap intervals.

\begin{table*}
\caption{Electron densities inferred from the [S II] doublet at
$T_e=10^4$ K. Central values are measured from the unresampled full-sample
median stacks; intervals are the 16th--84th percentiles of 100
galaxy-bootstrap realisations and are used only as uncertainties.}
\label{tab:measurements}
\centering
\begin{tabular}{llrrr}
\hline
Matching & Redshift & Low $\Sigma_\star$ & Middle $\Sigma_\star$ & High $\Sigma_\star$\\
 & & \multicolumn{3}{c}{$n_e$ (cm$^{-3}$)}\\
\hline
$M_\star$ & 0.10--0.20 & $32.2^{+1.9}_{-1.9}$ & $46.3^{+1.2}_{-1.9}$ & $79.7^{+1.5}_{-1.6}$\\
$N_{\rm stack}$ & & 122,236 & 122,236 & 122,236\\
$M_\star$ & 0.20--0.30 & $60.1^{+1.5}_{-1.6}$ & $85.3^{+1.9}_{-1.9}$ & $123.8^{+1.5}_{-2.0}$\\
$N_{\rm stack}$ & & 127,463 & 127,463 & 127,463\\
$M_\star$ & 0.30--0.44 & $91.2^{+2.3}_{-2.5}$ & $119.7^{+2.6}_{-2.6}$ & $149.0^{+2.6}_{-2.4}$\\
$N_{\rm stack}$ & & 113,379 & 113,379 & 113,379\\
$M_\star+\mathrm{SFR}$ & 0.10--0.20 & $32.3^{+2.0}_{-1.0}$ & $48.2^{+1.3}_{-1.3}$ & $70.3^{+1.3}_{-1.2}$\\
$N_{\rm stack}$ & & 116,272 & 116,272 & 116,272\\
$M_\star+\mathrm{SFR}$ & 0.20--0.30 & $62.0^{+2.3}_{-1.8}$ & $85.7^{+2.0}_{-1.9}$ & $118.2^{+1.7}_{-1.8}$\\
$N_{\rm stack}$ & & 121,519 & 121,519 & 121,519\\
$M_\star+\mathrm{SFR}$ & 0.30--0.44 & $99.8^{+2.6}_{-3.3}$ & $121.9^{+2.7}_{-2.7}$ & $145.9^{+2.0}_{-2.7}$\\
$N_{\rm stack}$ & & 107,846 & 107,846 & 107,846\\
\hline
\end{tabular}
\end{table*}

\section{Robustness and discussion}

\begin{figure*}
\centering
\includegraphics[width=0.98\textwidth]{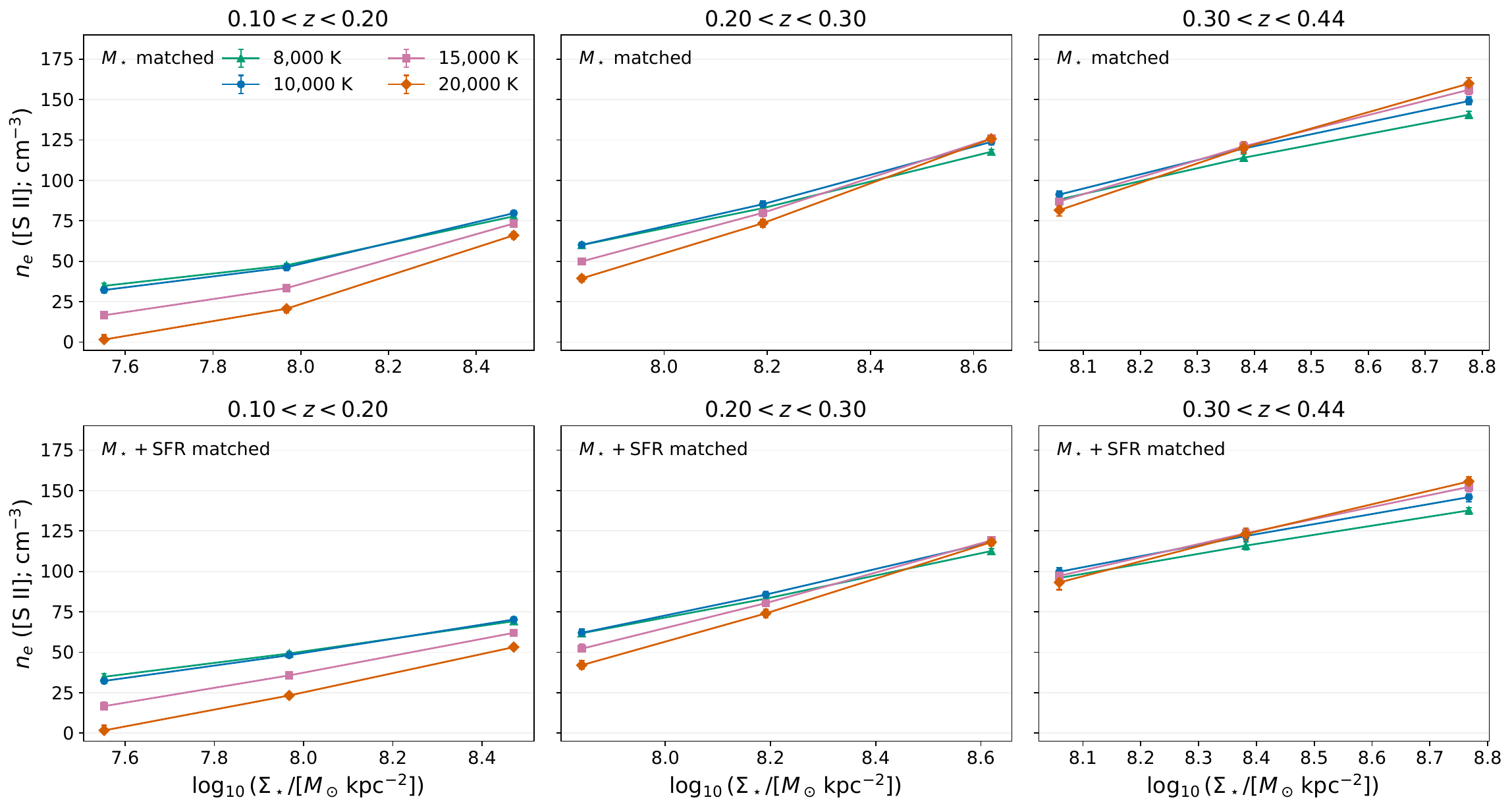}
\caption{Dependence of the $n_e$--$\Sigma_\star$ sequences on
the adopted electron temperature. Columns show redshift intervals; the upper
and lower rows show $M_\star$ and $M_\star+$SFR matching, respectively.}
\label{fig:temperature}
\end{figure*}

It is worth noting that the result in this work is a population relation, not a geometric identity. At fixed
$M_\star$ and SFR, stellar compactness may trace gas pressure, the spatial
concentration of star formation, or the relative contribution of compact H\,{\sc ii}
regions and diffuse ionized gas. Because every galaxy is continuum-normalized
and receives one equal vote in the fiducial median stack, the resulting $n_e$
is the average electron density of the galaxy population in each
bin. It is not a resolved pc-scale measurement of an individual H\,{\sc ii}
region.
The
decrease in the contrast after SFR matching indicates that part of the
relation is shared with the star-formation budget, while the remaining signal
shows that compactness contains additional information.  In physical terms,
the result is consistent with the idea that the kpc-scale stellar distribution
influences the gas pressure and confinement of smaller star-forming structures, but it does not
demonstrate a direct causal transfer of density from kpc to parsec scales.

Several physical mechanisms could produce this cross-scale connection.  First,
at fixed $M_\star$, a higher $\Sigma_\star$ implies a more concentrated stellar
potential.  Together with the gas surface density, this can raise the mid-plane
pressure and help confine denser star-forming clouds; pressure estimates based
on the combined stellar and gas surface densities provide a useful framework
for this expectation \citep{elmegreen1989,blitz2006,leroy2008}.  Second, total
SFR does not fix the area over which star formation occurs.  More compact
galaxies can therefore have a higher SFR surface density and a different
distribution of ionizing sources, even after their integrated SFRs are matched.
This possibility is consistent with the established connection between compact
galaxy structure, central mass build-up, and the concentration of star-forming
activity \citep{fang2013,barro2013,woo2015}.  Finally, changing
compactness may alter the relative contribution of classical H\,{\sc ii}
regions and diffuse ionized gas within the fibre.  Spatially resolved studies
show that diffuse ionized gas has distinct low-ionization-line properties and
can vary substantially across galaxies \citep{belfiore2016,lacerda2018}; a
change in this mixture could modify the integrated [S II] ratio without
requiring every individual H\,{\sc ii} region to follow the same density law.
These explanations are not mutually exclusive, and the present unresolved
stacks cannot distinguish among them.  Spatially resolved spectroscopy that
combines [S II] with independent pressure, ionization-parameter, gas-surface-
density, and diffuse-gas diagnostics will be needed to determine which process
dominates.  Such observations, together with resolved measurements of the SFR
surface density, offer a direct test of whether the relation is set primarily
by gas confinement, the concentration of star formation, or aperture-integrated
mixing.

The conversion from the doublet ratio to density depends on the adopted
electron temperature.  We use $T_e=10^4$ K as a fiducial value, but repeat the
conversion for $T_e=8000$, 15,000, and 20,000 K using the same measured
doublet ratios. Figure~\ref{fig:temperature} displays all three compactness
tertiles for the full set of assumed temperatures. The absolute value of
$n_e$ and the amplitude of the contrast vary, as expected because the [S II]
emissivity ratio is temperature dependent. Nevertheless, all tested
conversions give finite density solutions and preserve the monotonic ordering:
the high-compactness stack remains denser than the low-compactness stack in
every redshift interval and for both matching schemes. Thus the
$n_e$--$\Sigma_\star$ trend is insensitive to the fiducial temperature adopted
for the conversion.
It is therefore difficult to reproduce the observed $n_e$--$\Sigma_\star$
sequence through temperature variations alone.  Across the deliberately broad
$8000$--$20\,000$ K range, the temperature-induced change in $n_e$ for an
individual stack is at most $33.2^{+0.3}_{-1.0}$ cm$^{-3}$, smaller than the
minimum fiducial low-to-high-$\Sigma_\star$ density difference of
$38.0^{+1.9}_{-2.4}$ cm$^{-3}$ among our matched samples.  These intervals
are propagated from the same 100 galaxy-level bootstrap realisations.
Moreover, attributing the density sequence to
temperature would require an additional systematic trend in which $T_e$
decreases with increasing $\Sigma_\star$.  To our knowledge, such an inverse
$T_e$--$\Sigma_\star$ relation has not been reported observationally.

We also tested whether the inferred densities depend on the adopted stellar
population library.  We repeated the continuum subtraction with the FSPS SSP
templates \citep{conroy2009,conroy2010} distributed in the official
\textsc{pPXF} SPS-model repository, using the same age interval, fitting
window, emission-line masks, additive-polynomial degree, and common
instrumental resolution as in the eMILES analysis.  In both cases the
continuum-subtracted \sii\ doublet was measured with the same two-Gaussian
model.  The FSPS measurements show the same increasing
$n_e$--$\Sigma_\star$ trend as the fiducial eMILES measurements in every
redshift interval, both for the $M_\star$-matched sample and after the
additional SFR matching (see Fig.~\ref{fig:density}). Thus the central result is not specific to the
adopted SPS template library.

This distinction is important for interpreting the novelty of the measurement.
The individual ingredients are familiar: forbidden-line ratios measure
nebular density, and compactness is a well-established descriptor of galaxy
structure.  The new empirical step is to test their cross-scale covariance in
the same galaxies without requiring a detectable doublet in every object, and
then to ask whether it survives controls for the dominant global variables.
The matched median stacks turn that question into a measurement of a parent
population rather than a relation defined only by the brightest H\,{\sc ii}
regions.

The absence of a \sii\ S/N cut is important.  It allows the stacks to include
the weak-line population that cannot yield a stable individual density.  A
strong-line-only catalogue would preferentially select high-surface-brightness
H\,{\sc ii} regions and would not measure the average density of the parent
population.  The present analysis therefore turns the DESI sample size into a
measurement of the average cross-scale trend.  This choice follows the same
logic demonstrated by stacked JWST/NIRSpec density measurements
\citep{liurong2026}, while extending the experiment to a much larger DESI
sample and directly testing how the average nebular density changes across
galaxies with different kpc-scale stellar surface densities, after controlling
for stellar mass and SFR. Imposing a
doublet S/N threshold would also change the distributions of both
$\Sigma_\star$ and the measured $n_e$, because detectability depends on line
surface brightness, redshift, SFR, and the density-sensitive line ratio.
The resulting strong-line-only $n_e$--$\Sigma_\star$ relation would therefore
be incomplete and potentially biased toward bright, compact nebular regions,
rather than a reliable relation for the parent galaxy population.

As a secondary feature, we find that the characteristic densities of all three
compactness tertiles appear to increase toward the higher-redshift intervals.
This qualitative impression is consistent with previous reports of elevated
nebular density and interstellar-medium pressure at intermediate and high
redshift \citep[e.g.]{liurong2026,shirazi2014,sanders2016,kaasinen2017,kaasinen2018},
and could plausibly reflect the changing gas and star-formation conditions of
earlier galaxies.  We do not interpret this as a measurement of isolated
density evolution: a dedicated analysis would need to match $M_\star$ and SFR
across, rather than only within, the redshift intervals, and to examine the
changing physical area sampled by the DESI fibre.  Such an analysis is beyond
the main aim of this paper.  Our controlled result is the ordering with
$\Sigma_\star$ within each redshift interval.

\section{Conclusions}

We have used DESI DR1 \sii\ spectral stacking to test whether the average
nebular electron density, $n_e$, is correlated with kpc-scale galaxy stellar
surface density (compactness), $\Sigma_\star$.
In AGN-clean star-forming galaxies at $0.10<z<0.44$, the \sii\ ratio decreases
from low to high $\Sigma_\star$ in all three redshift intervals.  The inferred
$n_e$ increasing with increasing $\Sigma_\star$ relation remains unchanged
after matching $M_\star$ and after additionally matching SFR. These results
establish an empirical $n_e$--$\Sigma_\star$ relation connecting galaxy
compactness to the average electron density of ionized gas measured from
equal-galaxy-weight spectral stacks.

\begin{acknowledgements}
This study uses data from DESI.  We thank Hu Zou in NAOC for providing the DESI
stellar-mass and emission-line catalogue.  YR acknowledges support from the
CAS Pioneer Hundred Talents Program (Category B), NSFC grants 12522302, 12673017 and
12273037, and the USTC Research Funds of the Double-First-Class Initiative.
This work is supported by the China Manned Space Program with grant Nos.
CMS-CSST-2025-A06 and CMS-CSST-2025-A08. We used ChatGPT 5.6 Sol to polish the paper.
\end{acknowledgements}

\end{document}